\documentclass[showpacs,aps,twocolumn,superscriptaddress,prb,amsmath]
{revtex4}
\usepackage{txfonts}
\usepackage{color}
\usepackage[latin9]{inputenc}
\usepackage{amssymb}
\usepackage{graphicx}
\usepackage{dcolumn}
\usepackage{bm}
\usepackage{hyperref}
\usepackage{color}

\begin{document}
\title{New structure candidates for the experimentally synthesized heptazine-based and triazine-based two dimensional graphitic carbon nitride}
\author{Luneng zhao$^\sharp$}
\affiliation{Hunan Key Laboratory of Micro-Nano Energy Materials and Devices, Xiangtan University, Hunan 411105, P. R. China}
\affiliation{Laboratory for Quantum Engineering and Micro-Nano Energy Technology and School of Physics and Optoelectronics, Xiangtan University, Hunan 411105, P. R. China}
\author{Xizhi Shi$^\sharp$}
\affiliation{Hunan Key Laboratory of Micro-Nano Energy Materials and Devices, Xiangtan University, Hunan 411105, P. R. China}
\affiliation{Laboratory for Quantum Engineering and Micro-Nano Energy Technology and School of Physics and Optoelectronics, Xiangtan University, Hunan 411105, P. R. China}
\author{Jin Li}
\email{lijin@xtu.edu.cn}
\affiliation{Hunan Key Laboratory of Micro-Nano Energy Materials and Devices, Xiangtan University, Hunan 411105, P. R. China}
\affiliation{Laboratory for Quantum Engineering and Micro-Nano Energy Technology and School of Physics and Optoelectronics, Xiangtan University, Hunan 411105, P. R. China}
\author{Tao Ouyang}
\affiliation{Hunan Key Laboratory of Micro-Nano Energy Materials and Devices, Xiangtan University, Hunan 411105, P. R. China}
\affiliation{Laboratory for Quantum Engineering and Micro-Nano Energy Technology and School of Physics and Optoelectronics, Xiangtan University, Hunan 411105, P. R. China}
\author{Chunxiao Zhang  }
\affiliation{Hunan Key Laboratory of Micro-Nano Energy Materials and Devices, Xiangtan University, Hunan 411105, P. R. China}
\affiliation{Laboratory for Quantum Engineering and Micro-Nano Energy Technology and School of Physics and Optoelectronics, Xiangtan University, Hunan 411105, P. R. China}
\author{Chao Tang}
\affiliation{Hunan Key Laboratory of Micro-Nano Energy Materials and Devices, Xiangtan University, Hunan 411105, P. R. China}
\affiliation{Laboratory for Quantum Engineering and Micro-Nano Energy Technology and School of Physics and Optoelectronics, Xiangtan University, Hunan 411105, P. R. China}
\author{Chaoyu He}
\email{hechaoyu@xtu.edu.cn}
\affiliation{Hunan Key Laboratory of Micro-Nano Energy Materials and Devices, Xiangtan University, Hunan 411105, P. R. China}
\affiliation{Laboratory for Quantum Engineering and Micro-Nano Energy Technology and School of Physics and Optoelectronics, Xiangtan University, Hunan 411105, P. R. China}
\author{Jianxin Zhong}
\affiliation{Hunan Key Laboratory of Micro-Nano Energy Materials and Devices, Xiangtan University, Hunan 411105, P. R. China}
\affiliation{Laboratory for Quantum Engineering and Micro-Nano Energy Technology and School of Physics and Optoelectronics, Xiangtan University, Hunan 411105, P. R. China}

\begin{abstract}
The widely used crystal structures for both heptazine-based and triazine-based two-dimensional (2D) graphitic carbon nitride (g-C$_3$N$_4$) are the flat P-6m2 configurations. However, the experimentally synthesized 2D g-C$_3$N$_4$ possess thickness ranging in 0.2-0.5 nm, indicating that the theoretically used flat P-6m2 configurations are not the correct ground states. In this work, we propose three new corrugated structures P321, P3m1 and Pca21 with energies of 66 (86), 77 (87) and 78 (89) meV/atom lower than that of the corresponding heptazine-based (triazine-based) g-C$_3$N$_4$ in flat P-6m2 configuration, respectively. These corrugated structures have very similar periodic patterns to the flat P-6m2 ones and they are difficult to be distinguished from each other according to their top-views. The optimized thicknesses of the three corrugated structures ranging in 1.347-3.142 {\AA} are in good agreement with the experimental results. The first-principles results show that these corrugated structural candidates are also semiconductors with band gaps slightly larger than those of the correspondingly flat P-6m2 ones. Furthermore, they possess also suitable band edge positions for sun-light-driven water-splitting at both $pH=0$ and $pH=7$ environments. Our results show that these three new structures are more promising candidates for the experimentally synthesized g-C$_3$N$_4$.
\end{abstract}

\maketitle

\section{Introduction}
The two-dimensional (2D) graphitic carbon nitride (g-C$_3$N$_4$) is a fascinating conjugated polymer widely used as metal-free photocatalyst for sun-light-driven water-splitting in the area of solar energy conversion and environment remediation\cite{s1, s2, s3}. It is reported that g-C$_3$N$_4$ possesses excellent photocatalytic activities exceeding than the widely used nitrogen-doped TiO$_2$ \cite{s4} and g-C$_3$N$_4$ have attracted widespread concern \cite{s1, jcc2}. Many experimental works \cite{s1, jcc13, jcc14, jcc15, jcc16} have successfully synthesized the single-layered g-C$_3$N$_4$ and confirmed that they are better than the corresponding three-dimensional (3D) counterparts for photocatalytic applications. However, the structural details of the 2D g-C$_3$N$_4$ are still unfixed. Previous literatures \cite{spl, s7, s8, s9} have reported that at least seven types of 2D g-C$_3$N$_4$ can be constructed and four of them are related to segments of triazine (C$_3$N$_3$) and heptazine (C$_6$N$_7$) \cite{s6, s5}, with both hexagonal (P-6m2) and orthogonal (Pmc21) assembling manners. The heptazine-based g-C$_3$N$_4$ in both P-6m2 and Pmc21 configurations possess relatively larger nitrogen pores in comparison with the triazine-based ones and they are more stable than the corresponding triazine-based systems in energy \cite{spl, s7, s8, s9}. In the past decades, the flat P-6m2 models with energies slightly smaller than the flat Pmc21 ones are considered as the ground states for both the triazine-based and heptazine-based g-C$_3$N$_4$ and they are widely-used as the structure candidates for the synthesized triazine-based \cite{s10} and heptazine-based \cite{s11,small1,small2} g-C$_3$N$_4$.

Based on the flat P-6m2 model, many theoretical works \cite{jcc21, jcc23, jcc27, jcc29, jcc31} have been performed to investigate the fundamental properties and potential applications of 2D g-C$_3$N$_4$. Most of the calculated properties based on such a flat P-6m2 model can match the experimental results well. However, there are obvious evidences to show that the experimentally synthesized g-C$_3$N$_4$ possess thickness in the range of 0.2-0.5 nm \cite{jcc13, jcc15}, indicating that the flat P-6m2 configurations may not the configurations observed in experiments. Therefore, it of great importance to identify the true configurations of the experimentally synthesized g-C$_3$N$_4$. In fact, some previous theoretical works have considered using the corrugated structures to do their researches \cite{jpcc2019, jcc36, jcc37, jcc38, jcc39}. Proper atomic corrugations can release energies to make the systems more stable than the exactly flat ones. However, these fragmented works are lack of in-depth understanding of the structural details for the experimentally synthesized 2D g-C$_3$N$_4$. We have also noticed that some low-energy corrugated configurations \cite{JMC2009, prbPicard} have been previously proposed for 3D g-C$_3$N$_4$, but these corrugated configurations have not been translated to 2D conditions. Therefore, to systematically investigate the possible corrugated configurations for both the triazine-based g-C$_3$N$_4$ and heptazine-based g-C$_3$N$_4$ are still interesting and necessary work to understand the structure of 2D g-C$_3$N$_4$.

In this work, a simple method is proposed to systematically search for corrugated configurations of 2D g-C$_3$N$_4$. We found many new corrugated configurations for the triazine-based and heptazine-based 2D g-C$_3$N$_4$, including three high-symmetry ones (P321, P3m1 and Pca21) with remarkable stabilities. They possess energies of about 78 (89), 77 (87) and 66 (86) meV/atom lower than that of the corresponding heptazine-based (triazine-based) one in flat P-6m2 configuration, respectively. These three new structures and the flat P-6m2 one possess very similar periodic patterns and they are difficult to be distinguished from each other according to their top-views. However, the calculated layer thicknesses of the three new corrugated configurations are in the range of 1.347-3.142 {\AA}. After consider the atomic covalent radius at both side (0.77 {\AA} for carbon and 0.75 {\AA} for nitrogen), these layer-thicknesses shift to 2.847-4.642 {\AA}, which are good agreement with the experimental results. Further more, the first-principles calculations indicate that they are also insulators with energy band gaps just slightly larger than that of the correspondingly flat P-6m2 one. Their band edges suggest that they are also potential materials for sun-light-driven water-splitting at both $pH=0$ and $pH=7$ environments, comparable to the widely used P-6m2 configuration. In view of their excellent stabilities, proper thicknesses and electronic properties, we suggest that these three new corrugated configurations are promising structural candidates for the experimentally synthesized 2D g-C$_3$N$_4$.
\section{Methodologies}
\indent The widely used P-6m2 structure of the triazine-based and heptazine-based g-C$_3$N$_4$ are isomorphic to a hypothetical binary network NX (same as the crystal structure of hexagonal boron nitride), in which X is a hypothetical super-atom for indicating triazine or heptazine, and N indicates the nitrogen atoms as shown in Fig. 1 (a). We notice that to search possible corrugated configurations for 2D g-C$_3$N$_4$ is equal to search up and down sequences of N atoms in the hypothetical NX networks. Such a task can be quickly performed by our previously developed RG$^2$ code \cite{yhcprb, Shi18, hcyprl}. RG$^2$ code is a high-efficient code for generating crystal structures with well-defined structural-feature \cite{splpb, jnpss, lzqass, znprb, yxjap} and it can quickly provide us many inequivalent NX networks with different up and down sequences of N atoms, with only six-member rings. Associated by an additional small program, the X atoms in these hypothetical NX networks are replaced by triazine or heptazine to construct real corrugated structures for g-C$_3$N$_4$. Finally, the structures and stabilities of a series corrugated g-C$_3$N$_4$ structures with atoms no more than 56 are further investigated by first-principles calculations. The low-energy P321, P3m1 and Pca21 are further investigated by the high-level HSE06 method on their electronic properties.

\begin{figure*}
\begin{center}
\includegraphics[width=\textwidth]{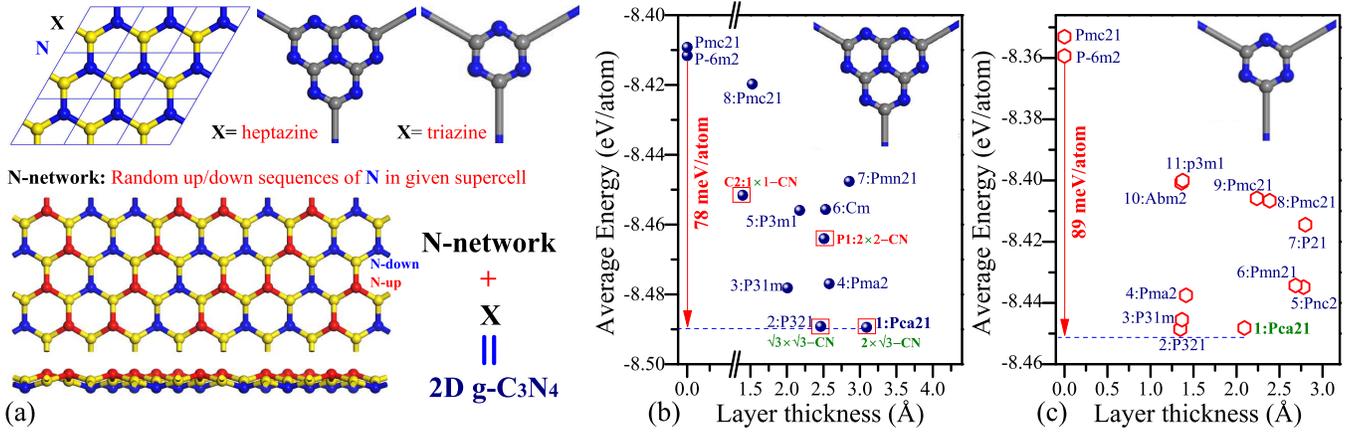}
\caption{(a) Sketch map for constructing corrugated configurations of 2D g-C$_3$N$_4$. Heptazine and triazine are considered as a three-coordinated super-atoms (X) to insert in the given three-connected binary network NX with different up and down sequences of N atoms, where N indicates the nitrogen atoms. (b) and (c) are the calculated average energies and layer thicknesses of the corrugated configurations for the heptazine-based and triazine-based 2D g-C$_3$N$_4$, respectively. For heptazine-based ones, the recently proposed $1$$\times$$1$-CN, $\sqrt{3}$$\times$$\sqrt{3}$-CN, $2$$\times$$2$-CN and $2$$\times$$\sqrt{3}$-CN are denoted by red squares.}
\end{center}
\end{figure*}

All calculations about structure optimization and property investigations are performed by the density functional theory based first-principles methods as implemented in the widely-used VASP code\cite{VASP}. The interactions between nucleus and the corresponding valence electrons are described by the projector augmented wave methods (PAW) \cite{PAW1, PAW2}. Interactions between valence electrons and the exchange-correlation energies are considered through the generalized gradient approximation (GGA) developed by Perdew, Burke, and Ernzerhof (PBE) \cite{PBE}. A plane wave basis with cutoff energy of 500 eV is used to expand the wave functions for all carbon systems and the Brillouin zone sampling meshes are set to be dense enough to ensure the convergence. All the 2D systems are fully optimized until the residual forces on every atom is less than 0.001 eV/${\AA}$. The convergence criterion of total energy is set to be 10$^{-7}$ eV and the thickness of the slab-model is set to be larger than 15 ${\AA}$ to avoid spurious interactions between adjacent images in our calculations. The used sample K-meshes are set to be denser-enough (Ka=40/a, Kb=40/b, Kc=1) to guarantee the accuracy. The high-level HSE06 \cite{HSE06} method is considered to investigate the electronic properties of the three low-energy configurations of g-C$_3$N$_4$.
\begin{figure*}
\begin{center}
\includegraphics[width=12cm]{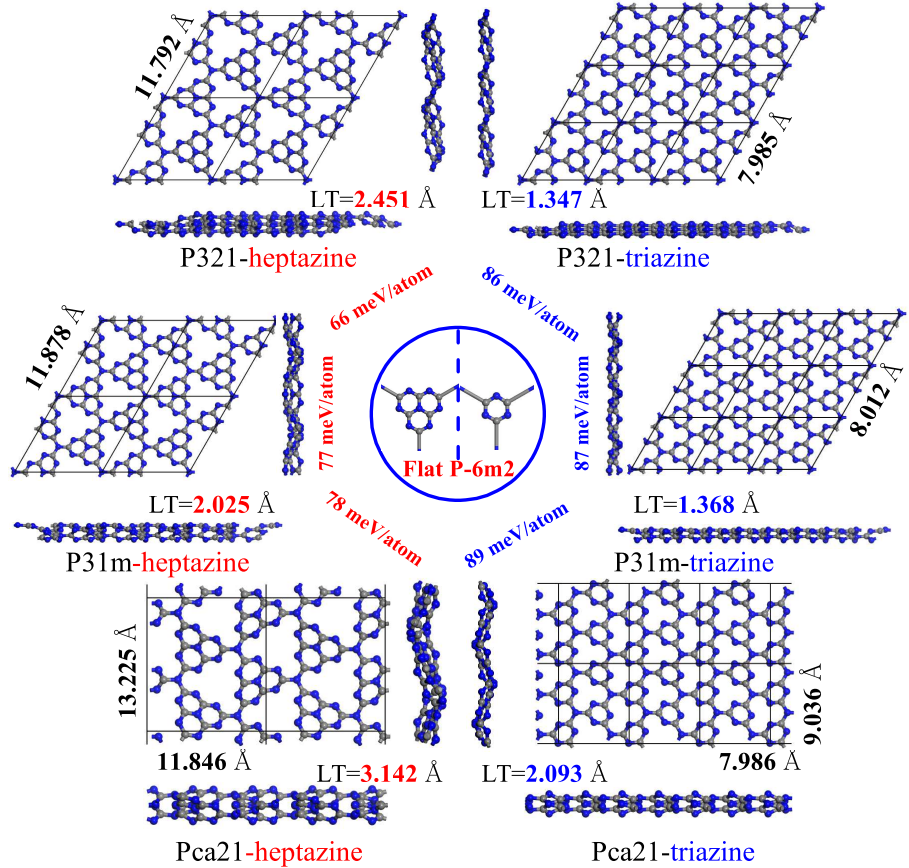}
\caption{The optimized crystal structures of the corrugated P321, P3m1 and Pac21 for both heptazine-based (left) and triazine-based (right) g-C$_3$N$_4$ shown in top and side views with their corresponding lattice constants, layer-thicknesses (LT) and the released energies.}
\end{center}
\end{figure*}
\section{Results and discussions}
\subsection{Structures and stabilities}
As shown in Fig. 1 (b), we find 10 corrugated configurations for the heptazine-based 2D g-C$_3$N$_4$, including the recently proposed $2$$\times$$\sqrt{3}$-CN (P321) and $2$$\times$$2$-CN (Pca21) \cite{newjpcc}. In the recent work\cite{newjpcc}, there are 4 corrugated configurations have been proposed for heptazine-based 2D g-C$_3$N$_4$, including $1$$\times$$1$-CN and $2$$\times$$2$-CN which are missed in our present work. The calculated average energies shown in Fig. 1 (b) suggest that these corrugated configurations are energetically more stable than the corresponding flat P-6m2 one. For the triazine-based ones, there are 12 corrugated configurations as shown in Fig. 1 (c). We notice that all these configurations have not been discussed for the triazine-based 2D g-C$_3$N$_4$ in the recent work\cite{newjpcc}. Their calculated energies are also lower than that of the widely-used flat P-6m2 configuration. These results suggest that the widely-used P-6m2 configurations are not the ground states for both the heptazine-based and triazine-based 2D g-C$_3$N$_4$.

There are three high-symmetry structures showing excellent energetic stabilities in Fig.1 (b) and (c) for both triazine-based and heptazine-based g-C$_3$N$_4$. We record them as P321, P3m1 and Pca21 according to their hexagonal and orthogonal symmetries, respectively. As shown in Fig. 2, the corrugated P321, P3m1 and Pca21 are of about 78, 77 and 66 meV/atom lower than that of the corresponding heptazine-based P-6m2 configuration, respectively. For the triazine-based ones, the released energies will be further increased to be 89, 88 and 86 meV/atom, respectively. These remarkable released energies can be qualitatively understood by the corrugations. The using of z-direction leaves the system enough space to arrange their atoms and enlarge the distance between unpaired electrons in the systems. Thus, the coulomb interactions in the system will be weaken and this will correspondingly enhance the stabilities of the systems. That is to say, the flat P-6m2 configuration is hard to exist in nature. Such a point has been further confirmed by the recent work according to their calculated vibrational spectrum\cite{newjpcc}. The flat P-6m2 is dynamically unstable according to the imaginary frequency in its phonon band structure. And such imaginary frequencies can be eliminated through proper structural corrugations ($1$$\times$$1$-CN, $\sqrt{3}$$\times$$\sqrt{3}$-CN, $2$$\times$$2$-CN and $2$$\times$$\sqrt{3}$-CN)\cite{newjpcc}.

To study the structural features of these three new corrugated structures, we show their optimized crystal structures in Fig.2 in both top and side views. We notice that P321 and P3m1 are two different reconstructions of the hexagonal $\sqrt{3}$$\times$$\sqrt{3}$ supercell ($\frac{\sqrt{3},0,0}{0,\sqrt{3},0}$) of the flat P-6m2, and Pca21 is another reconstruction of P-6m2 in its orthogonal supercell ($\frac{2,1,0}{0,2,0}$), as indicated in Fig. S1. They all can be directly discovered by RG$^2$ code in graph-associated normal search process or phase-transition process. In Fig. 2, the optimized lattice-constants and layer thicknesses of the three low-energy configurations are also provided. We can see that the lattice constants of the optimized P321 (a=b=11.792 {\AA} for heptazine-based or 7.985 {\AA} for triazine-based), P3m1 (a=b=11.878 {\AA} for heptazine-based or 8.012 {\AA} for triazine-based) and Pca21 (a=11.846 {\AA} and b=13.225 {\AA} for heptazine-based or a=7.986 {\AA} and b=9.036 {\AA} for triazine-based) are very close to those of the corresponding supercells based on the flat P-6m2 model as shown in Fig. S1. For the heptazine-based ones, our results for P321 and Pca21 are very close to those reported in recent work\cite{newjpcc}, which confirmed that the calculation method and settings in our present work are correct. Especially, the top views of these three corrugated configurations are nearly same as those of the flat P-6m2 structures, showing very similar periodic patterns. That is to say, it is difficult to distinguish these structures in experiment from their top views. This might be an important reason that previous researchers consider using the flat P-6m2 model as the structural candidate for understanding the experimentally synthesized g-C$_3$N$_4$.

The widely-used used P-6m2 model is exactly flat without thickness. However, there are some previous experiments suggested that the synthesized single-layer of g-C$_3$N$_4$ possess obvious thicknesses about 2-5 {\AA} \cite{jcc13, jcc15}. In Fig. 2, we can see that the three low-energy configurations P321, P3m1 and Pca21 are obvious corrugated. For the heptazine-based 2D g-C$_3$N$_4$, the layer thicknesses are 2.451, 2.025 and 3.142 {\AA}, respectively. And these layer thickness will reduce to 1.347, 1.368 and 2.093 {\AA} for the triazine-based ones. After consider the atomic covalent radius of carbon (0.77 {\AA}) and nitrogen (0.75 {\AA}) at both side, these layer-thicknesses ranged in 2.847-4.642 {\AA} are proper for explaining the experimental observations \cite{jcc13, jcc15}.

We have also checked the effect of dispersion interactions in these systems. As shown in Table S1, we can see that the dispersion interactions just slightly reduce the lattice constants of these configurations. The layer-thicknesses are also slightly enlarged due to the dispersion interactions. Especially, the stability order for these singly-layer g-C$_3$N$_4$ are not changed due to the included dispersion interactions. These results provide a fact that lattice constants and layer thicknesses are mainly decided by the strong inner-layer covalent C-C and C-N bonds. That is to say, the weak inner-layer dispersion interactions in such strong covalent single-layer systems can be ignored.
\subsection{Electronic properties}
We then investigate the electronic properties of these three low-energy g-C$_3$N$_4$ through the high-level HSE06 method and compare them with those of the widely used flat P-6m2. The calculated band structures of both the triazine-based and heptazine-based g-C$_3$N$_4$ in the flat P-6m2 and the corrugated P321, P3m1 and Pac21 configurations are shown in Fig. S2. The results show that the heptazine-based g-C$_3$N$_4$ in flat P-6m2 is a semiconductor with indirect band gap of 2.77 eV and the triazine-based one possesses direct band gap of about 3.184 eV. These results are good agreement with those calculated in previous theoretical works \cite{spl, jcc21, jcc23, jcc27}, which confirm that our used methods are reliable.
\begin{figure}
\begin{center}
\includegraphics[width=\columnwidth]{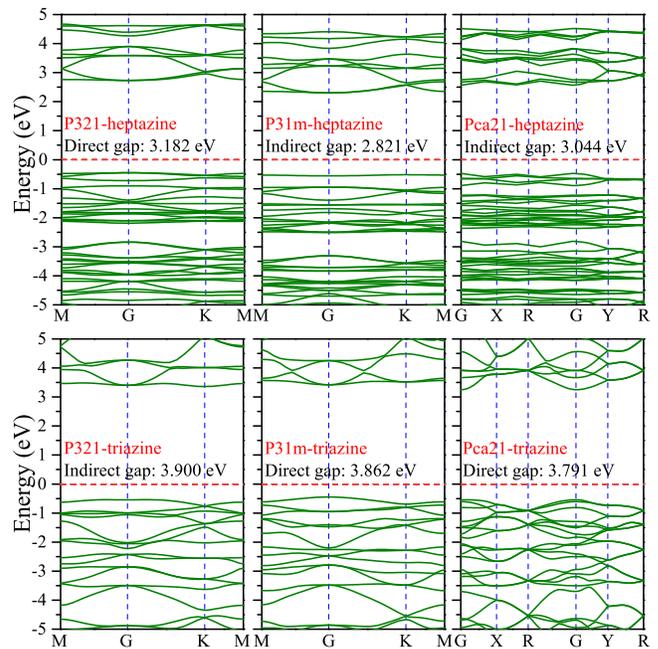}
\caption{The electron band structures of the triazine-based and heptazine-based g-C$_3$N$_4$ in configurations the corrugated P321, P3m1 and Pac21 calculated from HSE06 methods. The corresponding band gaps and types are also inserted.}
\end{center}
\end{figure}
\begin{figure}
\begin{center}
\includegraphics[width=\columnwidth]{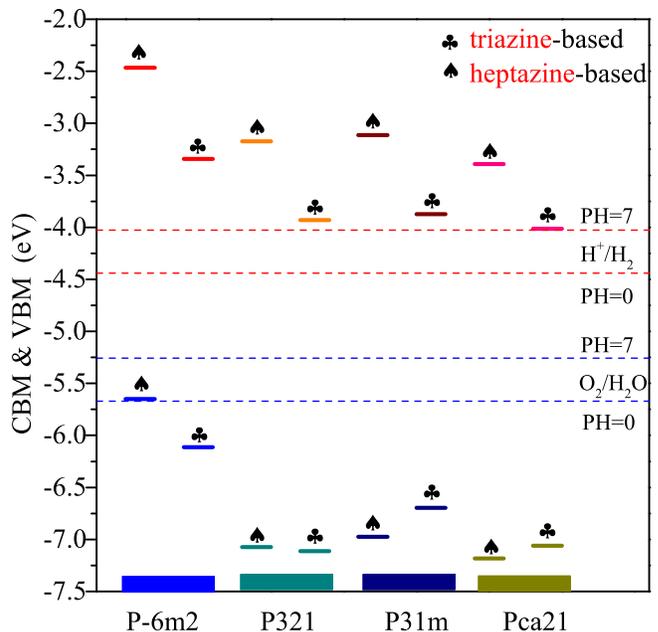}
\caption{The HSE06-based CBM and VBM positions of the triazine-based and heptazine-based g-C$_3$N$_4$ in configurations of flat P-6m2 and corrugated P321, P3m1 and Pac21 with respect to the redox potentials for water splitting at $pH=0$ and $pH=7$ situations.}
\end{center}
\end{figure}

As discussed before, proper atomic corrugations will release corresponding energies to make the systems more energetically stable. Here, we can see that these corrugations (P321, Pm31 and Pca21) will also enlarge the band gaps of the systems to enhance their chemical stabilities. As the band structures of the heptazine-based g-C$_3$N$_4$ shown in Fig. 4, we can see that the splitting energies between the valence bands and conducting bands are enlarged due to the atomic corrugations. The band gaps of the heptazine-based g-C$_3$N$_4$ in the corrugated P321, P3m1 and Pac21 are correspondingly increased to be 3.182 eV, 2.821 eV and 3.044 eV, respectively. For the triazine-based g-C$_3$N$_4$, the band gaps of the corrugated P321, P3m1 and Pac21 are also enlarged to be 3.90 eV, 3.862 eV and 3.791 eV, respectively. These increased splitting energies or band gaps between the conduction bands and valence bands can help us to partially understand the corrugation-induced energy release (the electron part) in the systems. For the heptazine-based g-C$_3$N$_4$, it is found that the amplifications of band gaps between flat and corrugated g-C$_3$N$_4$ (P321, Pm31, Pca21 are 0.412 eV, 0.051 eV and 0.274 eV, respectively) are relatively small. That is to say, the calculated band gaps of heptazine-based g-C$_3$N$_4$ in flat P-6m2 and corrugated P321, Pm31 and Pca21 are very close to those reported in previous experiments as shown in Table S2 (2.89 eV \cite{jcc13}, 2.95 eV \cite{K4} and 2.92 eV \cite{K5}). Thus, previous literatures can use the calculated band gap based on the flat P-6m2 model to explain the experimental values well, no matter the experimentally synthesized g-C$_3$N$_4$ is the flat P-6m2 or the corrugated P321, Pm31 and Pca21. For the triazine-based g-C$_3$N$_4$, these amplifications are 0.716 eV, 0.678 eV and 0.607 eV for P321, Pm31 and Pca21, respectively, which indicate that it is easy to know if the experimentally synthesized g-C$_3$N$_4$ is flat or corrugated. However, it is difficult to find experimental band gap for the single layer of triazine-based g-C$_3$N$_4$ at present time \cite{s10}.

Another important feature of the experimentally synthesized g-C$_3$N$_4$ is the exotic properties for metal-free photocatalysis. Based on the flat P-6m2 models, the excellent potentials of g-C$_3$N$_4$ for sun-light-driven water-splitting have been well understood by their CBM and VBM positions satisfying the redox potentials of the water splitting \cite{spl, jcc27, jcc37, jcc39, jpcc2019}. We think this is another important reason why the flat P-6m2 models are considered as structure candidates for the experimentally synthesized g-C$_3$N$_4$. Here, we show that the low-energy P321, P3m1 and Pac21 are also excellent materials for metal-free sun-light-driven water-splitting. As shown in Fig.5, the HSE06-based CBM and VBM positions of the triazine-based and heptazine-based g-C$_3$N$_4$ in configurations of flat P-6m2 and corrugated P321, P3m1 and Pac21 are arranged together with the redox potentials for water splitting at $pH=0$ and $pH=7$ situations. For both $pH=0$ and $pH=7$ situations, all the CBM positions of the corrugated configurations located slightly above the redox potential of H$^+$/H$_2$. And the corresponding VBM positions are also lower than the redox potential of O$_2$/H$_2$O. These band edge positions of the three new corrugated configurations suggest that they are also appropriate for sun-light-driven water splitting in both acidic and neutral situations.

Above results and discussions can help us to understand why the flat P-6m2 models are widely used as the crystal structures for the experimentally synthesized g-C$_3$N$_4$ in previous literatures. We think that the most important reason is that there is no any optional structure to use. The widely-used flat P-6m2 models can be accepted as the correct structures due to the facts that their surface topographies (top views), electronic properties (band gaps) and potentials in sun-light-driven water-splitting (band edge positions) are very similar to those of the experimentally synthesized corrugated structures. It is still difficult to say what is the true configurations for the corrugated 2D g-C$_3$N$_4$ at present time. Maybe the structure morphologies of the synthesized 2D g-C$_3$N$_4$ are substrate-dependent. The three new corrugated configurations are promising candidates in views of their excellent stabilities, proper thicknesses and electronic properties. We believe that the orthogonal Pca21 configurations are the most possible candidates according to their lowest energies. We have also show that they are dynamically stable according to the simulated vibrational spectrum as shown in Fig. S3. Such dynamical stabilities these corrugated configurations of 2D g-C$_3$N$_4$ have also be confirmed by the recent theoretical work\cite{newjpcc}.
\section{Conclusion}
In summary, three corrugated configurations P321, P3m1 and Pca21 were proposed as new candidate structures for the experimentally synthesized g-C$_3$N$_4$. The first-principles calculations show that these three new configurations are more energetically favorable than the flat P-6m2 models. These three corrugated configurations possess similar periodic patterns with those of flat P-6m2 models. However, the corrugated P321, P3m1 and Pca21 with certain layer-thicknesses are more suitable for explaining the thicknesses of the experimentally synthesized g-C$_3$N$_4$. The calculated band gaps of the corrugated P321, P3m1 and Pca21 are also close to those calculated based on the flat P-6m2, which can further help us to understand the previous belief of using the P-6m2 model. Furthermore, the band edge positions of these three new corrugated configurations for both triazine-based and heptazine-based g-C$_3$N$_4$ are also appropriate for sun-light-driven water-splitting in both acidic and neutral situations. Our results show that these three new corrugated configurations are promising structure candidates for the experimentally synthesized single layer of g-C$_3$N$_4$.

\section{accknowledgement}
This work is supported by the National Natural Science Foundation of China (Grants No. 11974300, 11974299, 11704319 and 11874316), the Natural Science Foundation of Hunan Province, China (Grant No. 2016JJ3118 and 2019JJ50577), and the Program for Changjiang Scholars and Innovative Research Team in University (No. IRT13093).

$\#$ These authors contributed equally to this work.
\bibliographystyle{apsrev}

\end{document}